\documentclass[9pt,twocolumn,twoside]{osajnl}

\journal{ao} 

\setboolean{shortarticle}{false}
\usepackage{blindtext}
\usepackage{color,soul}

\title{Spectropolarimeter on-board the Aditya-L1: Polarization Modulation and Demodulation}

\author[*]{K. Nagaraju}
\author{B. Raghavendra Prasad}
\author{Bhavana S. Hegde}
\author{Suresh Venkata Narra}
\author{D. Utkarsha}
\author{Amit Kumar} 
\author{Jagdev Singh}
\author{Varun Kumar}
\affil[]{Indian Institute of Astrophysics, Koramangala Second Block, Sarjapur Road, Bengaluru-560034, India\\
}

\affil[*]{Corresponding author: nagarajuk@iiap.res.in}

\setul{0.5ex}{0.3ex}
\definecolor{Red}{rgb}{1,0.0,0.0}
\setulcolor{Red}

\begin{abstract}
One of the major science goals of the Visible Emission Line Coronagraph (VELC) payload aboard the Aditya-L1 mission is to map the coronal magnetic field topology and the quantitative estimation of longitudinal magnetic field on routine basis.
The infrared (IR) channel of VELC is equipped with a polarimeter to carry out full Stokes spectropolarimetric observations in the Fe~{\sc xiii} line at 1074.7~nm .
The polarimeter is in dual-beam setup with continuously rotating waveplate as the polarization modulator.
Detection of circular polarization due to Zeeman effect and depolarization of linear polarization in the presence of magnetic field due to saturated Hanle effect in the Fe~{\sc xiii} line require high signal-to-noise ratio (SNR).
Due to limited number of photons, long integration times are expected
to build the required SNR.
In other words signal from a large number of modulation cycles are to
be averaged to achieve the required SNR.
This poses several difficulties. One of them is the increase in data
volume and the other one is the change in modulation matrix in
successive modulation cycles. The latter effect arises due to a 
mismatch between the retarder's rotation period and the length of the
signal detection time in the case of VELC spectropolarimeter  (VELC/SP).
It is shown in this paper that by appropriately choosing the number of
samples per half rotation the data volume can be optimized.
A potential solution is suggested to account for modulation matrix
variation from one cycle to the other. 
\end{abstract}

\setboolean{displaycopyright}{true}

\begin{document}

\maketitle

\section{Introduction}
Aditya-L1 is the first Indian space mission to study the Sun \cite{Singh2011,RaghavendraPrasad2017}. 
One of the major observational goals of the Aditya-L1 mission is to carry out direct measurement of coronal magnetic fields on synoptic basis.
VELC is the largest among the seven payloads aboard the Aditya-L1 aimed towards studying the corona in emission lines based on coronagraphic technique.
This payload has a continuum channel to record the coronagraphic
images in the continuum with the central wavelength close to
530~nm covering the field-of-view (FOV) extending up to 3~$R_\odot$ ($R_\odot$ - the radius of the Sun) and three spectroscopy channels. The spectrograph is designed to record
coronal spectra simultaneously in three spectral lines at 530.3~nm
due to Fe~{\sc xiv} (green channel), 789.2~nm due to Fe~{\sc xi} (red channel) and 1074.7~nm due to Fe~{\sc xiii} (IR channel). 
This is a multi-slit spectrograph (4 equispaced slits) covering the FOV up to $1.5R_\odot$ \cite{Rajkumar2018}.
The starting FOV for all the channels is $1.05R_\odot$.
The IR channel has a polarimeter to carry out full Stokes spectropolarimetric observations of the Fe~{\sc xiii} line.

The polarimeter of VELC consists of a continuously rotating waveplate (RWP) as the polarization modulator followed by a polarizing
beam displacer (PBD) as the polarization analyser.  
The PBD produces two beams which are orthogonally polarized which
are simultaneously recorded by the detector, 
the setup commonly known as dual-beam polarimetry \cite{Donati90, Semel93}.
One of the main advantages of the dual-beam setup is to minimize the spurious polarization i.e. cross-talk from Stokes~$I$ to $Q$, $U$ and $V$ due to intensity fluctuations caused for e.g. by the satellite jitter in the case of VELC. However, the rotation frequency of the VELC modulator which is about 0.1~Hz is much above the expected frequency of the satellite jitter and hence the jitter induced spurious polarization expected is low. In this case the signals from the individual channels can be demodulated separately to get the corresponding Stokes parameters and later averaged to increase the SNR by a factor of about $\sqrt{2}$. However, in the scenario where the frequency of the satellite jitter is larger than the modulation frequency then one may have to take the difference signal between the two channels and then carry out demodulation to obtain the Stokes parameters. In this case precision in flat fielding will decide the final precision achieved in the polarization measurements. Based on the initial measurements on-board one may decide on which of the two approaches mentioned above is suitable. 

Rotating waveplate is a relatively simple, versatile and robust polarization modulator with a single optical component using which full Stokes polarimetry can be carried out \cite{Lites87}. 
In this system the duty cycle that can be achieved is as high as $100\%$.
The input polarization signal gets modulated in such a way that the Stokes~$Q$ and $U$ have 4 times but with a phase difference of 90 degrees and Stokes~$V$ has 2 times the retarder rotation frequency.
The modulation frequency is independent of the retardance of the retarder. 
However, the amplitudes and the Stokes $Q$ offset depend on the retardance value.
In order to have maximum efficiency for Stokes~$V$ a quarter-wave plate (QWP)
is chosen for VELC polarimetry which will be apparent in the following section.
With the choice of QWP as the modulator the Stokes $Q$ offset is 0.5 and hence the modulated Stokes Q signal stays always above zero, the effect
of which is discussed in the later part of the paper.
Another feature of the continuously rotating waveplate is that there
exists numerous signal sampling schemes. 
Of many possibilities, in its original study Lites \cite{Lites87}
has adapted eight equally spaced samplings of the modulated signal per half rotation of the waveplate. Following this almost all the
polarimeters built based on the continuously RWP adapted the detection 
scheme of eight samples per half rotation (for e.g., Advanced Stokes Polarimeter \cite{Elmore92}, SPINOR \cite{Socas-Navarro06}, Chromospheric Lyman-Alpha Spectro-Polarimeter\cite{Ishikawa2011}) though some polarimeters such as Hinode/SP have adapted 8 as well as 4 samples per
half rotation\cite{Ichimoto2008}. Such a flexibility is very useful in the context of
VELC/SP. Exploring an optimum number of samples per half rotation by taking in to account various observational constraints and the hardware 
limitations forms part of the research presented in this paper.

The number of samples per half rotation $n$ dictates the exposure time
per detector frame which is given by $T/2n$, where $T$ is the 
retarder rotation period. However, the actual detector frame time ($\Delta\tau$)
could differ from that of $T/2n$. In order to distinguish these two
times we introduce a terminology retarder slot time defined as 
\begin{equation}
    \Delta t = \frac{T}{2n}.
    \label{eq:Deltat}
\end{equation}

Polarimetric observations in general are photon starved and the scarcity
is even more severe in coronal spectropolarimetric observations \cite{Lin2000,Lin2004}. 
Hence signal from several modulation cycles are to be averaged to achieve the required SNR. The required SNR to achieve the science goals envisaged by Aditya-L1/VELC is in the range 10$^2$ (in Stokes Q and U measurements \cite{Gibson2017}) to 10$^4$ (in Stokes-V). Given the expected number of photons seen by the
VELC detectors \cite{Singh2019}, one is required to integrate the signal over several minutes to hours amounting to a large number of modulation cycles for one single measurement. There are two aspects of major concern to VELC/SP due to long integration measurements.
One is the data volume which increase with increase in integration time
as there is no onboard data processing is planned. 
The second aspect is, even if there is a minute difference between
$\Delta t$ and $\Delta\tau$ the modulation matrix changes in
successive modulation cycles. This problem is specific to VELC/SP or any
other polarimeter based on RWP which do not have a hardware handshake between the detector and the polarization modulator.

Various aspects of the source of errors in polarization measurements have
been studied by different authors such as misalignment, scale error,
uncertainty in the retardance and etc \cite{Ishikawa2011,Liang2019}.
While these sources of errors are to be kept at minimum, the main focus
of this paper is to formulate, analyze and find a potential solution to
address the change in modulation matrix from one modulation cycle to
another caused due to non-zero difference between $\Delta t$ and 
$\Delta\tau$. The aim is to find a solution which does not require
any hardware change because in space astronomy modification in hardware
is an involved process.

\section{SNR Requirement and Integration Times}

The Fe~{\sc xiii} at 1074.7~nm being a magnetic transition (forbidden) line can provide information only on the line-of-sight (LOS) magnetic field strength from circular polarization i.e. Stokes~$V$ profiles  and field orientation in the plane-of-the-sky (POS) through linear polarization i.e. Stokes~$Q$ and $U$ profiles \cite{Sahal-Brechot1977, LinCasini2000}.  
Due to large Doppler broadening the weak field approximation (WFA) works well for the IR line to derive the LOS component of the magnetic field \cite{Lin2000, Lin2004}.
Under WFA the Stokes $V$ signal is directly proportional to the LOS field strength and the derivative of Stokes $I$ w.r.t. wavelength as given by 
the following equation 
\begin{equation}
    V(\lambda) = K B \frac{dI(\lambda)}{d\lambda},
    \label{eq:WFA}
\end{equation}
where $K = 6.5\times 10^{-6}$ nmG$^{-1}$ for the IR line. 
In order to estimate Stokes $V$ as a function of field strength the only unknown in Eq.~\ref{eq:WFA} is the first derivative of the intensity profile. 
This is not a readily available quantity in the literature though one can get a rough estimation of this quantity from published magnetic field measurements carried out using this line for e.g. by Lin et. al. \cite{Lin2000,Lin2004}.
In order to estimate the first derivative of Stokes $I$ we have used the data from Norikura observatory. The spectroscopic data used in this
analysis are part of the observations carried out using the
25~cm coronagraph of the Norikura observatory by Singh et al. \cite{Singh1999,Singh2002,Singh2003}.
The spectral resolution of these ground based observations which is $\approx 0.006$~nm is much better than $\approx 0.05$~nm that is expected to be achieved with the VELC. Hence, the selection of these data for estimating the intensity derivative is justified.

\begin{figure}[htbp]
\centering
\fbox{\includegraphics[width=\linewidth]{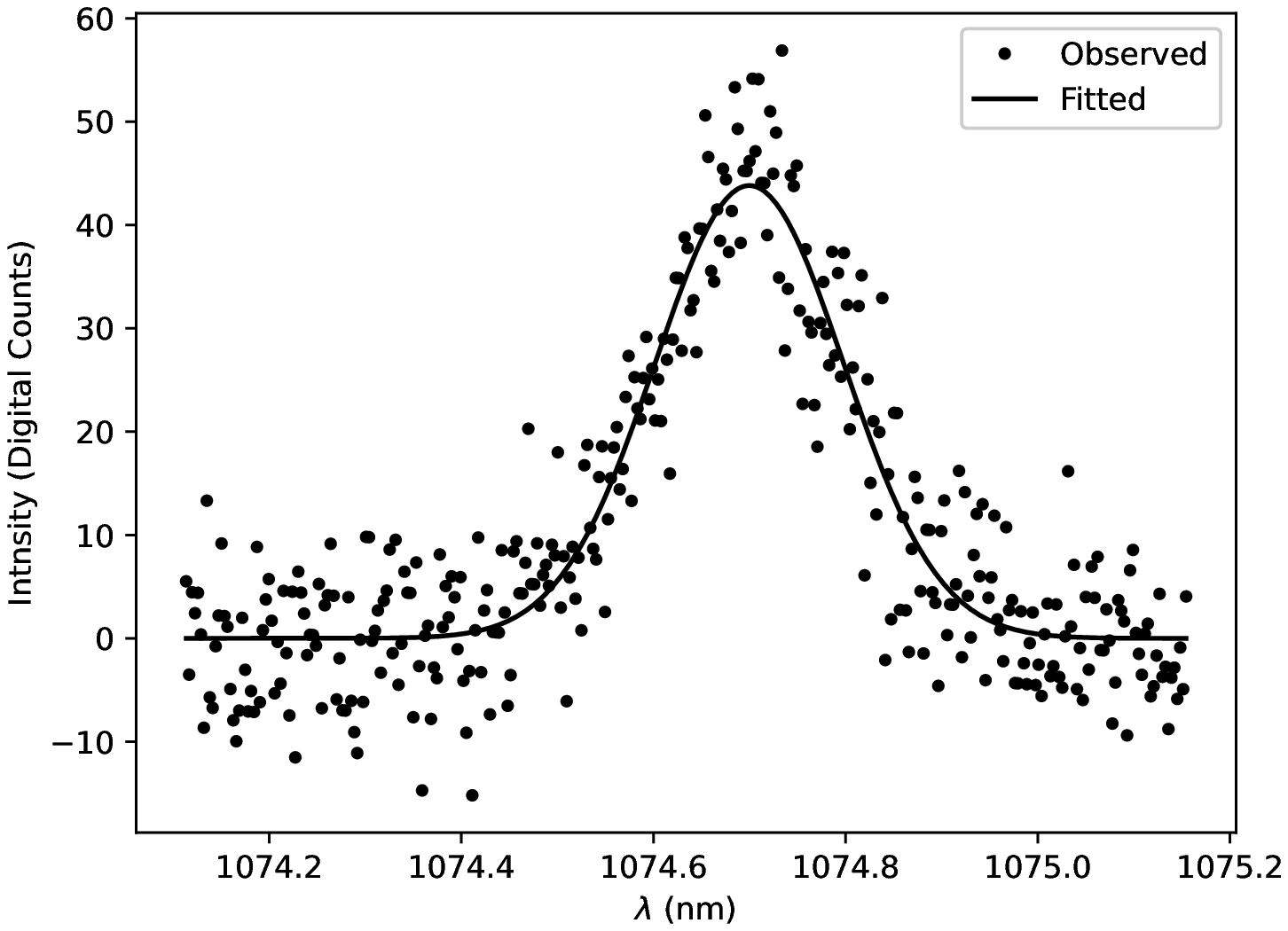}}
\caption{A sample observed profile of Fe~{\sc xiii} at 1074.7~nm (the dotted curve) and a Gaussian fitted profile (the solid curve).}
\label{fig:profgaussfit}
\end{figure}

\begin{figure}[htbp]
    \centering
    \includegraphics[width=\linewidth]{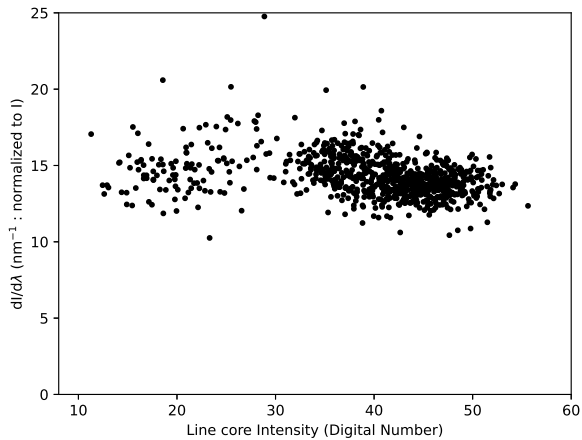}
    \caption{Scatter plot of first derivative of the intensity profile close to FWHM normalised to local intensity value and the line core intensity (nothing but the fitted Gaussian amplitude)} in digital counts.
    \label{fig:plotintgrad}
\end{figure}

\begin{figure}[htbp]
    \centering
    \includegraphics[width=\linewidth]{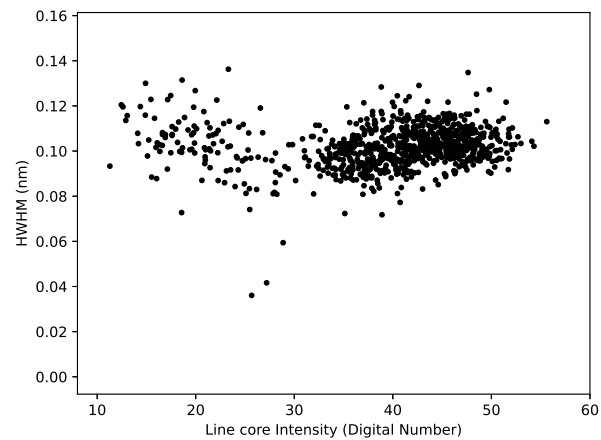}
    \caption{Scatter plot of HWHM of the intensity profile and the line core intensity in digital counts.}
    \label{fig:plotHWHM}
\end{figure}

The spectral profiles are fitted using a Gaussian function. While fitting\ul{,} the slope in the observed profile is taken in to account by fitting a straight line along with the Gaussian. A sample observed profile after correcting for the continuum slope (the dotted curve) and a fitted Gaussian profile (the solid curve) are shown in Fig.~\ref{fig:profgaussfit}. 
The intensity derivative with respect to wavelength is calculated from
the fitted profile in order to keep the effect of observed noise to the minimum. 
The intensity derivative is expected to peak 
close to Full Width at Half Maximum (FWHM) and so the Stokes V amplitude. 
Hence, the derivative of the intensity profile close to FWHM is considered for estimating Stokes~$V$ amplitude.
A scatter plot of the intensity gradient close to FWHM and the line core
intensity (nothing but the fitted Gaussian amplitude) is shown in Fig.~\ref{fig:plotintgrad}. 
This representation is chosen because the coronal emission line
intensity distribution covers the regions of coronal loops, 
outside of the loops as well as different coronal heights.
There are two clusters of points below and above the line core intensity
of about 30 (digital counts).
For the lower intensity cluster the intensity derivative shows increasing trend with increasing line core intensity where as the trend is opposite for the higher intensity cluster.
The average intensity derivative for a given line core intensity ranges
between 13 and 16 as shown in the scatter plot.
The corresponding expected Stokes~$V$ signal for a 1~Gauss LOS field is $8.5 \times 10^{-5}$ and $1 \times 10^{-4}$, respectively (cf. Eq.~\ref{eq:WFA}).
These expected Stokes~$V$ amplitudes are about 4-5 times higher than the measurements reported by Lin et al. \cite{Lin2004}.
According to Eq.~\ref{eq:WFA} this can happen if the first derivative of Stokes~$I$ observed by Lin et al. \cite{Lin2004} is smaller by that factor. 
In order to cross-check our estimations presented in this paper we have used another method described by Tomczyk et al. \cite{Tomczyk2008} which
relates the error in magnetic field estimation with that of error in Stokes~$V$ measurement. 
According to Tomczyk et al. \cite{Tomczyk2008} the error in magnetic field estimation ($\sigma_B$) is related to error in Stokes~$V$ measurement ($\sigma_V$) through

\begin{equation}
    \sigma_B = 87398~ w~ \sigma_V,
    \label{eq:BerrHWHM}
\end{equation}
where $w$ is the half-width-at-half-maximum (HWHM) of the emission line in nm.
A scatter plot of HWHM and the line core intensity of the IR is shown in Fig.~\ref{fig:plotHWHM}. 
Here also there are two clusters of points can be seen located above and below the line core intensity of about 30 (digital counts). 
The average HWHM at a given line core intensity decreases from about 0.12~nm to 0.085~nm with increase in line core intensity. 
And the corresponding error in Stokes~$V$ according to Eq.~\ref{eq:BerrHWHM} are $9.5\times10^{-5}$ and $1.3\times10^{-4}$, respectively which causes an error of 1~Gauss in the LOS field estimation.
For the higher intensity cluster the average HWHM increases from about
0.095~nm to about 0.108~nm with increase in line core intensity.
And the corresponding error in Stokes~$V$ is $1.2\times10^{-4}$ and $1\times10^{-4}$ which leads to LOS field estimation with an error of 1~Gauss.
These estimations are consistent with the above estimations based on the first derivative of Stokes~$I$ (cf. Eq.~\ref{eq:WFA}). 

Given the limited number of photons available in coronal emission
line observations, a large number of measurements have to be
averaged to achieve a precision level of $10^{-4}$ or better.
We make use of the photon numbers estimated by Singh et al. \cite{Singh2019} to estimate the total integration time to achieve the 
required SNR in VELC spectropolarimetry. 
The Table~\ref{tab:SNR1} lists the number of photons and
the corresponding total integration time required to achieve
a SNR of $10^4$ per pixel at different coronal heights expressed in solar
radius $R_\odot$. 
The number of photons is given for the line core but the SNR and the integration time are estimated for the wavelength at FWHM.
The SNR per pixel per second is calculated by assuming only the photon
noise. 
The integration times will increase once the detector dark noise, 
readout noise and other sources of noise are taken in to account.
The numbers presented in the Table~\ref{tab:SNR1} are sufficient to
understand the issues discussed in the following section.

\begin{table}[htbp]
    \centering
    \caption{The number of photons ($N_p$) and the integration time ($T_\mathrm{int}$) in seconds required to achieve a SNR of $10^4$ in the photon noise limited case at different coronal heights given in terms of solar radius ($R_\odot$). The number of photons corresponds to the peak of the emission line due to Fe~{\sc xiii} at 1074.7 nm where as the SNR and
    $T_\mathrm{int}$ correspond to the wavelength at FWHM.}
    \begin{tabular}{|c|c|c|c|}
        \hline
        $R_\odot$ & $N_p$ & SNR/pix/s & $T_{\mathrm{int}}(s)$ \\
                  &       &           & (SNR=$10^4$)\\
        \hline
        1.05 & 19820 & 141 & 5045\\ 
        1.10 & 14144 & 119 & 7070\\ 
        1.15 & 11324 & 106 & 8830\\ 
        1.20 & 8484 & 92 & 11786\\ 
        1.3 & 6224 & 79 & 16067\\ 
        1.4 & 3964 & 63 & 25227\\ 
        1.5 & 2260 & 48 & 44247\\ 
        \hline
    \end{tabular}
    \label{tab:SNR1}
\end{table}

\section{Polarization Modulation, Demodulation and Calibration}
As we have seen in the previous section that due to limited
number of photons long integration times are needed to achieve
the required SNR, particularly for Stokes $V$. This implies that
the polarization measurements from a large number of polarization
modulation cycles have to be averaged. 
In this context we discuss about two 
aspects specific to VELC payload viz. optimization of data volume
and the effect of difference between $\Delta t$ and $\Delta\tau$
on the polarimetric accuracy. 
In the following section we discuss the working principle of the RWP based modulator which is helpful in understanding these issues.

\subsection{Working Principle of Continuously Rotating Waveplate Based Modulator}

\begin{figure}[htbp]
    \centering
    \includegraphics[width=\linewidth]{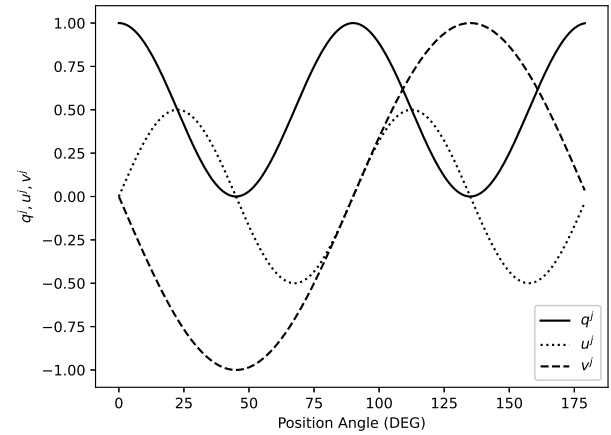}
    \caption{Plots of modulation of Stokes $Q$ (the solid curve), $U$ (the dotted curve) and $V$ (the dashed curve) parameters as a function of position angle (nothing but the orientation of the fast axis of the waveplate w.r.t Stokes~$Q$ direction).}
    \label{fig:PlotsModSig}
\end{figure}

In the case of continuously rotating waveplate based modulator 
the modulated intensity recorded by the detector as a function of input Stokes parameters is given by
\begin{equation}
    I_\mathrm{mod}(\theta) = I + (c_2^2+s_2^2 \mathrm{cos}\delta)~ Q + c_2 s_2 (1-\mathrm{cos}\delta) ~U - s_2 \mathrm{sin}\delta ~V,
    \label{eq:modint}
\end{equation}
where $c_2 = \mathrm{cos}(2\theta)$ and $s_2 = \mathrm{sin}(2\theta)$ and
$I$, $Q$, $U$ and $V$ are the input Stokes parameters with $I$ 
representing the total intensity, $Q$ and $U$ defining the linear polarization state and the plane of polarization and $V$ defining the circular polarization state of the incoming beam.
If $n$ is the number of equally spaced samples per half rotation
then the exposure time is $T/2n$, where $T$ is the rotation period
and the angle over which the signal is integrated is given by $\pi/n$. Eq.~\ref{eq:modint}  can be integrated to get 
the coefficients of $Q$, $U$ and $V$ for the $j^{\mathrm{th}}$ sampling step or
the modulation step as
 \begin{equation}
     I^j_{mod} = I + q^j Q + u^j U + v^j V,
      \label{eq:modint2}
 \end{equation}
where
\begin{eqnarray}
q^j & = & \frac{1}{\Delta\theta^j} \int_{\theta_1^j}^{\theta_2^j} (c_2^2+s_2^2 ~\mathrm{cos}\delta) ~ d\theta, \\\nonumber 
u^j & = & \frac{1}{\Delta\theta^j} \int_{\theta_1^j}^{\theta_2^j} c_2~s_2~ (1-\mathrm{cos}\delta) ~d\theta,   \\
v^j & = & -\frac{1}{\Delta\theta^j} \int_{\theta_1^j}^{\theta_2^j} s_2 ~\mathrm{sin}\delta ~d\theta,  \nonumber 
 \label{eq:modint3}
\end{eqnarray}
where $j=1,2...n$ and  $\Delta\theta^j = |\theta_2^j-\theta_1^j|$ is the range of angle spanned by the fast axis of the waveplate over a given exposure time: $\theta_1^j = (j-1) \Delta\theta$ and  $\theta_2^j = j \Delta\theta$. 
For e.g., in the sampling scheme of $n=8$ equally spaced intervals over half rotation of the waveplate, $\Delta\theta$ is equal to $22.^o5$. 
In the above equation it is assumed that the input Stokes 
parameters do not change within one modulation cycle which is
a valid assumption in the case of coronal spectropolarimetry.
Plots of the coefficients $q^j, u^j$ and $v^j$ as a function of
rotation angle are given Fig.~\ref{fig:PlotsModSig}. 
For generating these plots the waveplate considered is a QWP ($\delta = 0.25$). 
The Stokes~$Q$ coefficient $q^j$ stays always above zero because of the offset which is 0.5 for the case of QWP. 

The Eq.~\ref{eq:modint2} can be written in a vectorial form for $n$~number of modulation steps or the number of signal detections per half rotation as \cite{delToroIniesta00}

\begin{equation}
    {\bf I} = {\bf O} {\bf S},
    \label{eq:modintVec}
\end{equation}
where ${\bf I}$ is the vector of length $n$ containing the modulated intensities,
${\bf O}$-is the $n\times 4$ modulation matrix and $\bf{S}$ is the input Stokes vector. 
The input Stokes vector is retrieved from the measured modulated intensities through
\begin{equation}
    {\bf S} = {\bf D} {\bf I}.
        \label{eq:demod}
\end{equation}
Where ${\bf D}$ is called the demodulation matrix. The most optimum 
demodulation matrix is given by the Moore-Penrose pseudo inverse of ${\bf O}$ as \cite{delToroIniesta00}
\begin{equation}
    {\bf D} = ({\bf O}^T {\bf O})^{-1} {\bf O}^T.
    \label{eq:demodmatopt}
\end{equation}
And the corresponding optimum efficiencies are defined as \cite{delToroIniesta00}
\begin{equation}
    \epsilon_i = \left({n \sum_{j=1}^n D_{ij}^2} \right)^{-\frac{1}{2}},
    \label{eq:effopt}
\end{equation}
where $i=1..4$ corresponding to $I$, $Q$, $U$ and $V$.

The modulation matrix changes according to the detection scheme
adapted.
In its original study Lites \cite{Lites87} adapted a detection
scheme which samples the encoded $Q$, $U$ and $V$ signals symmetrically
about each of their respective peaks and the first exposure
is centered about the Stokes Q amplitude.
The corresponding modulation matrix for the case of VELC with 8 equally
spaced samples per half rotation is given by 
\begin{equation}
{\bf O} =
\begin{pmatrix}
1.0 & 0.95 & 0.0 & 0.00 \\
1.0 & 0.50 & 0.45 & -0.69\\
1.0 & 0.05 & 0.0 & -0.97\\
1.0 & 0.50 & -0.45& -0.69\\
1.0 & 0.95 & 0.00 & 0.00 \\
1.0 & 0.50 & 0.45 & 0.69\\
1.0 & 0.05 & 0.00 & 0.97\\
1.0 & 0.50 & -0.45 & 0.69
\end{pmatrix}
.
\end{equation}
As can be seen from the modulation matrix this is not a well balanced modulation scheme. A balanced scheme should have more or less equal weightage to Stokes $Q$, $U$ and $V$ in all modulation steps so that all the intensity measurements are utilised so that a better SNR is achieved.
The modulation can be better balanced if the exposure of the detector is synchronized to start with the retarder rotation. 
In that case the modulation matrix becomes
\begin{equation}
\label{eq:modmat8meas}
{\bf O} =
\begin{pmatrix}
1.0 & 0.82 & 0.32 &  -0.37 \\
1.0 & 0.18 &  0.32 &  -0.90 \\
1.0 & 0.18 & -0.32 & -0.90 \\
1.0 & 0.82 & -0.32 & -0.37 \\
1.0 & 0.82 & 0.32 &  0.37 \\
1.0 & 0.18 &  0.32 &  0.90 \\
1.0 & 0.18 & -0.32 & 0.90 \\
1.0 & 0.82 & -0.32 & 0.37 
\end{pmatrix}
.
\end{equation}

The detection schemes adapted later than Lites\cite{Lites87} have all adapted the above mentioned detection scheme, for example  Collados \cite{Collados07b} or CLASP \cite{Ishikawa2014}.
Most of the polarimeters based on continuously rotating waveplate
adapt the signal detection scheme in which there are 8 equally spaced samplings per half rotation of the waveplate with some exceptions like
in the case of Hinode/SP 4 samples per half rotation are also used \cite{Ichimoto2008}.
As noted by Lites \cite{Lites87} there are variety of detection schemes possible. This flexibility is particularly useful for the space based observatories such as Aditya-L1 
because the data download is limited due to limited telemetry. 
The data volume increases with increase in number of detector frames.
By reducing the number of samples per half rotation the data
volume can be reduced for the same total integration time.
For e.g., instead of 8 if 4 samples are chosen then the data 
volume will be reduced by 50\%. Similarly if 6 measurements
are chosen then the data volume is reduced by 25\%. 
Another advantage of reducing the number of samples is that the
overall readout noise will be less because the readout noise does not depend on the exposure time but defined per pixel per frame. 
However, the number of samples have to be chosen by taking in to account various other aspects of the measurements which are discussed in the following.

 \begin{table}[htbp]
\label{tab:effopt}
    \centering
    \caption{Optimum efficiencies of Stokes parameters as defined in Eq.~\ref{eq:demodmatopt} as a function of number of samples per half rotation ($n$) of the waveplate. While estimating the efficiencies the effect of non-zero difference of $\Delta t$ and $\Delta\tau$ is also taken in to account. The differences are tabulated in Table.~\ref{tab:SlotTime}. The values in brackets are when the difference is zero.}
    \begin{tabular}{|c|c|c|c|c|}
        \hline
        n  & \multicolumn{4}{c|}{Optimum Efficiencies} \\
        \cline{2-5}
        & $\epsilon_I$ & $\epsilon_Q$ & $\epsilon_U$ & $\epsilon_V$ \\
        \hline
        4 & 0.012 [0.000] & 0.006 [0.000] & 0.143 [0.269] & 0.640 [0.651] \\
        5 & 0.471 [0.472] & 0.265 [0.268] & 0.271 [0.268] & 0.664 [0.661] \\
        6 & 0.504 [0.505] & 0.296 [0.292] & 0.285 [0.292] & 0.668 [0.675] \\
        7 & 0.528 [0.524] & 0.303 [0.308] & 0.316 [0.308] & 0.694 [0.683] \\
        8 &  0.530 [0.537] & 0.323 [0.318] & 0.308 [0.318] & 0.675 [0.689] \\
        9 & 0.548 [0.546] & 0.323 [0.326] & 0.329 [0.326] & 0.698 [0.693] \\
        10 & 0.552 [0.552] & 0.330 [0.331] & 0.331 [0.331] & 0.696 [0.696] \\
        20 & 0.570 [0.571] & 0.348 [0.351] & 0.347 [0.348] & 0.703 [0.704] \\
        33 & 0.575 [0.576] & 0.352 [0.351] & 0.351 [0.351] & 0.705 [0.706] \\
        49 & 0.576 [0.576] & 0.353 [0.353] & 0.352 [0.353] & 0.706 [0.707] \\
        95 & 0.577 [0.577] & 0.353 [0.353] & 0.354 [0.353] & 0.707 [0.707] \\
        \hline
    \end{tabular}
    \label{tab:effSignalSamples}
\end{table}

The first thing is to check if there is any dependency of polarization
efficiency as a function number of samples. 
The Eq.~\ref{eq:effopt} is used to calculate the polarization 
efficiency as a function of different number of samples per half rotation.
The calculated efficiencies of the individual Stokes parameters are tabulated in Table.~\ref{tab:effSignalSamples}. 
These values are calculated by taking in to account the difference
between $\Delta t$ and $\Delta\tau$ as tabulated in Table.~\ref{tab:SlotTime}. 
The numbers inside the brackets are when there is no difference between
$\Delta t$ and $\Delta\tau$.
The efficiencies for the case of $n=4$ are weird because of the $q^j$s
are all equal to 0.5 with no sign change in all 4 measurements.
This causes ambiguity in delineating Stokes $I$ and $Q$ and hence no proper
demodulation matrix is available. 
However, if one works with the difference signal from the dual beam 
setup then Stokes $I$ should be equal to zero and it will be straight
forward to demodulate the Stokes parameters from the measured modulated
intensities.
The numbers in Table~\ref{tab:effSignalSamples} suggest that the efficiency of the Stokes parameters slowly increase with increase in number of samples per half rotation of the waveplate.
For Stokes $I$ and $U$ the increase is close to 10\% from the case of
$n=5$ to $n=95$, for Stokes $Q$ it is about 8\% and for Stokes $V$ it is about 4\%. This benefit of increase in efficiency will be offset by
the enormous increase in data volume as the number of samples increases from $n=5$ to $n=95$ (cf. Table~\ref{tab:SlotTime}). 
Hence the number of modulation steps has to be chosen by taking in to account the data volume and other measurement aspects.
For e.g., the starting FOV of VELC is 1.05$R_\odot$. The estimated photon count at this height is about 20000 per pixel per second (Singh et al. \cite{Singh2019} and cf. Table~\ref{tab:SNR1}). Since the IR camera is going to be operated in high gain mode for spectropolarimetric observations with a gain factor 7, the corresponding digital counts will be $\approx 2830$ per pixel per second. 
If we adapt a detection scheme of 4 samples per half rotation then the corresponding exposure time is 1.26~s as the rotation period of the retarder is 10.084~s. With this exposure time the digital counts expected per pixel is $\approx 3680$. This is too close to the maximum dynamic range of 12 bit ADC of the IR camera (see Sing et al. \cite{Singh2019} for more details on the detector). 
In a similar fashion various other aspects have to be taken in to account
before deciding on the number of samples per half rotation.
The Table~\ref{tab:SlotTime} provides closest frame time ($\Delta\tau$) of the detector to the retarder rotation slot time ($\Delta t$) as a function of different number of equally spaced samples per half rotation and the corresponding digital counts at the starting FOV of the VELC. The $6^\mathrm{th}$ column of this table provides the expected data volume per second.

\begin{table*}[ht]
    \centering
    \caption{Table of slot time ($\Delta t$), closest frame time ($\Delta\tau$) and their differences along with the maximum digital counts expected (close to $1.05 R_\odot$) per pixel for the case of different samples per half rotation of the wave plate. The values given in the square brackets in the fourth column are $n\times(\Delta\tau-\Delta t)$. The $6^\mathrm{th}$ column provides the expected data volume per second and the last column provides the minimum number of modulation cycles ($N_\mathrm{cyc}$)  for which the calibration data are required.}
    \begin{tabular}{|c|c|c|c|c|c|c|}
        \hline
        $n$ & $\Delta t$ & $\Delta\tau$ & $\Delta\tau-\Delta t$  & Digital counts & Data Volume & $N_\mathrm{cyc}$\\
         & (ms) &  (ms) & (ms) & (at 1.05 $R_\odot$) & per second (Mb)& (min)\\
        \hline
        4 & 1260.30  & 1253 & -7.30 [-29.2] & 3548 & 3.14 & 173 \\ 
        5 & 1008.24 & 1003 & -5.24 [-26.2] & 2840 & 3.92 & 194\\ 
        6 & 840.20 & 853 & 12.80 [76.8] & 2415 & 4.61 & 66\\ 
        7 & 720.17 & 703 & -17.17 [-120.2] & 1990 & 5.59 & 42\\ 
        8 & 630.15 & 653 & ~22.85 [182.8] & 1849 & 6.02 & 28\\ 
        9 & 560.13 & 553 & -7.13 [-64.17] & 1566 & 7.11 & 76\\ 
        10 & 504.12 & 503 & -1.12 [-11.2] & 1427& 7.81 & 450\\ 
        20 & 252.06 & 253 & 0.94 [18.8] & 716& 15.54 & 268 \\
        33 & 152.76 & 153 & 0.24 [7.92] & 433& 25.70 & 637\\
        49 & 102.88 & 103 & 0.12 [5.88] & 291& 38.17 & 858\\
        95 & 53.065 & 53 & -0.065 [-6.2] & 150& 74.19 & 816 \\
        \hline
    \end{tabular}
    \label{tab:SlotTime}
\end{table*} 
\subsection{Effect of retarder rotation slot time and the detector frame
time on the modulation}

\begin{figure}[htbp]
    \centering
    \includegraphics[width=\linewidth]{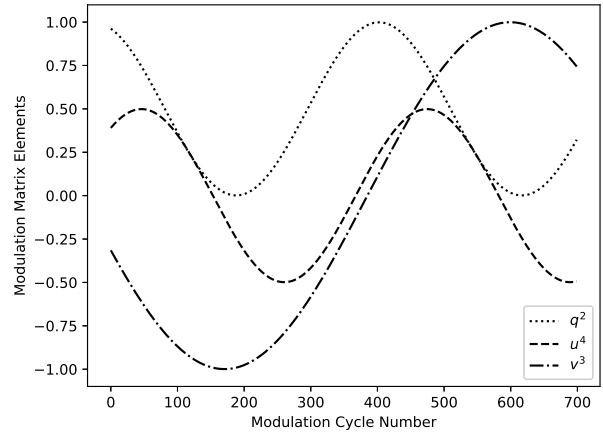}
    \caption{Plots of three randomly chosen modulation matrix elements corresponding to the detection scheme of $n=49$ as a function of cycle number. The variation is due to difference in rotating waveplate slot time and detector frame time of 0.12~ms.}
    \label{fig:plotsquv3ele49}
\end{figure}

 The detector that is going to be used for SP observations has the frame time which is either 53~ms or $53+m \times 50$~ms or integral multiples of
 these through frame binning, where $m$ is an integer. 
 On the other hand the retarder rotation period  is $10.084 $~s which is fixed. Given this it is not possible to precisely match the
 detector frame time ($\Delta\tau$) and the retarder rotation slot time ($\Delta t$). 
 The retarder slot time as a function of number of samples per half
 rotation ($n$) is given in the second column of Table~\ref{tab:SlotTime}. 
 The corresponding frame time that is closest to the slot time is listed in the third column of this table. 
 The difference in the slot time and the detector frame time is given in the fourth column. The fifth column provides the digital counts at $1.05~R_\odot$ for the corresponding  detector frame time and the expected data volume per second in the sixth column. 
 After the case of $n=10$ we have considered only a few selected cases
 for which the frame time and slot time are close by. 
 As can be seen from the table that the lowest difference between the
 slot time and the frame time is found with the frame time of 53~ms.
 However, what is more important is the difference between
 the time required for the retarder to complete $180^o$ rotation and
 the total time taken by the detector to record required number of samples for one modulation cycle i.e. $n\times(\Delta\tau-\Delta t)$. These differences are given inside the brackets of the fourth column of Table~\ref{tab:SlotTime}. 
 This can be understood as follows.
 For a fixed rotation of the waveplate the span of the angle sweeped by the fast axis of the retarder is determined by the detector frame time in the signal detection process. 
 Accordingly the coefficients of Stokes parameters as given
 in Eq.~\ref{eq:modint3} are determined and the corresponding modulation matrix is constructed.
 As long as the integral multiple of the detector's frame time is equal
 to the retarder's rotation period, i.e. $n \times\Delta T = n \times \Delta\tau$, then there are no issues whether
 the measurements require just one modulation cycle or the signals
 are to be integrated over several modulation cycles.
 When $n \times\Delta T \ne n \times \Delta\tau$ the fast axis position or the zero reference angle will be different when the signal sampling begins for successive modulation cycles. Hence the elements of the  modulation matrix start to change from one cycle to the other.  How fast the values of the modulation  matrix change depend on
 the magnitude of the difference.
 Even if the difference is small the effect can become significant
 in the measurements which require long integration due to accumulation
 effect.
 As an example, plots of 3 randomly chosen modulation matrix elements as a function of  modulation cycle number corresponding to the case when $n=49$ (with  the lowest time difference) are shown in Fig.~\ref{fig:plotsquv3ele49}. 
 The total number of modulation cycles considered here are 720
 corresponding to an integration time of one hour. 
 Such integration times are expected in order to detect
 Stokes $V$ signal produced due to Zeeman effect in the IR line (cf. Table~\ref{tab:SNR1}).
  
\begin{figure}[htbp]
    \centering
    \includegraphics[width=\linewidth]{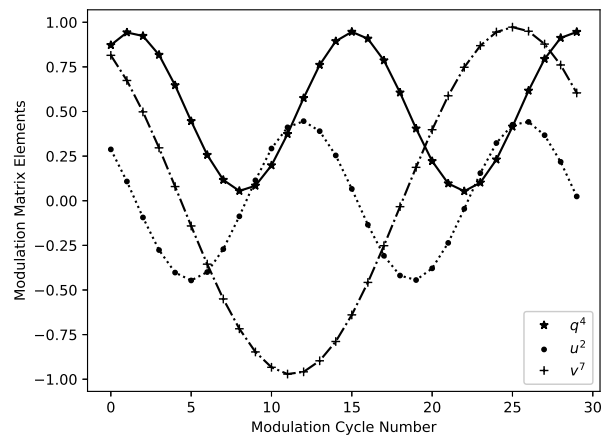}
    \caption{Same as Fig.~\ref{fig:plotsquv3ele49} but for the detection scheme with $n=8$ and different modulation matrix elements.}
    \label{fig:plotsquv3ele8}
\end{figure}

\begin{figure}[ht]
    \centering
    \includegraphics[width=\linewidth]{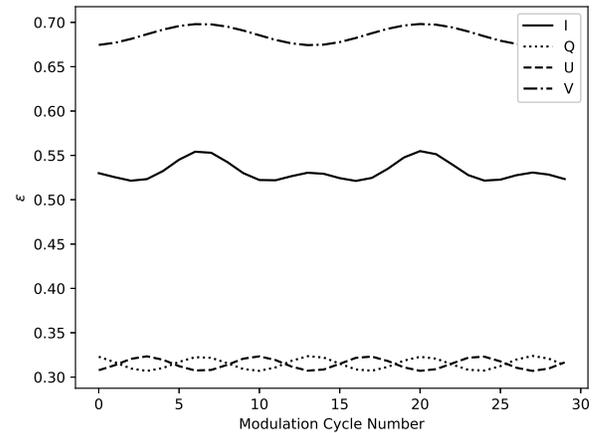}
    \caption{Plots optimum efficiencies for Stokes parameters as a function of modulation cycle number for the case of 8 samples per half rotation. }
    \label{fig:plotseffdemod}
\end{figure}

 The numbers in Table~\ref{tab:SlotTime} suggest that the absolute
 difference between slot time and the frame time decreases with increase
 in the number of samples per half rotation of the waveplate. 
 As number of samples increase the data volume also increases.
 Moreover, as suggested by the plots in Fig.~\ref{fig:plotsquv3ele49}, however small the difference is, due to accumulation effect, the modulation matrix changes significantly after a certain number of
 modulation cycles. If the difference is large then the modulation matrix
 differs from one modulation cycle to the next itself. 
 As a demonstration, plots of 3 randomly chosen modulation matrix element as a function of modulation cycle number are shown in 
 Fig.~\ref{fig:plotsquv3ele8} for the signal detection scheme of $n=8$
 which has the maximum difference between $\Delta\tau$ and $\Delta t$ among the cases discussed in this paper.
 The number of modulation cycles considered are only 30 as there is already a periodicity is seen in the variation of
 modulation matrix elements. This happens because of the faster variation of modulation matrix elements due  to larger time difference. 
 In spite of significant variation in individual elements of the modulation matrix the efficiency of Stokes parameters do not significantly change as a function modulation cycle number. 
 As an example, plots of efficiencies as a function of modulation cycle number for the case of $n=8$ are shown in Fig.~\ref{fig:plotseffdemod}.
 
 The above exercise suggests that even if the difference between
 the slot time and the frame time is small, in long integration
 measurements, the modulation matrix is expected to change in 
 successive modulation cycles.
 This will not become an issue if there is a handshake between
 retarder rotation and recording of the data by the detector such that
 the data recording is triggered only when retarder fast axis crosses the
 reference angle (in the case of VELC it is zero degree w.r.t. the polarization analyser's transmission axis). 
 In the case of VELC/SP there is no handshake exists between
 the retarder rotation and the detection process.
 In addition there is no mechanical shutter which means the
 pixels are always exposed to the light and mixing up of signals
 from different position angles is expected even if there is a handshake.
 If the data are recorded continuously then there is no mixing of
 signals happen but the modulation matrix changes in successive
 modulation cycles. Fortunately this change of modulation matrix can
 be taken care through polarimetric calibration which is explained
 as follows.
 
 \subsection{Calibration}
Standard polarimetric calibration can be adapted here as well. Since the
modulation matrix elements vary as a function of cycle number, the calibration data should be obtained for several modulation cycles. 
The number of modulation cycles ($N_\mathrm{cyc}$) for calibration should be equal to or more than the number of modulation cycles over which the signal is going to be integrated to achieve the required SNR. 
However, as can be seen in  Figs.~\ref{fig:plotsquv3ele49}~and~\ref{fig:plotsquv3ele8}, the modulation matrix elements vary periodically as a function of modulation cycle number. The modulation matrix elements $q^j$s and $u^j$s vary two times faster than $v^j$s.  
Hence, recording the calibration data for the number of modulation cycles within which the $v^j$s complete one period of variation may suffice. 
The last column of Table~\ref{tab:SlotTime} provides the minimum number of modulation cycles for which the calibration data are required as a function of different number of samples per half rotation. 
Larger the difference between the slot time and the detector frame time smaller the minimum $N_{cal}$ needed as the modulation matrix elements vary faster for higher difference. 


Let $N$ be the number of input known Stokes vectors and ${\bf S}_\mathrm{cal}$ be the
$4 \times N $ matrix containing the $N$ input known Stokes vectors.
And ${\bf I}_\mathrm{mod}$ be the corresponding $n \times N$ signal matrix with $n$ being the number of samples per half rotation. These are related by (cf. Eq. \ref{eq:modintVec})
\begin{equation}
    {\bf I}_\mathrm{mod} = {\bf O} {\bf S}_\mathrm{cal}.
\end{equation}
The above equation can be solved for the modulation matrix as 
\begin{equation}
    {\bf O} = \left( {\bf I}^T_\mathrm{mod} {\bf I}_\mathrm{mod} \right)^{-1} {\bf I}^T_\mathrm{mod} ~ {\bf S}_\mathrm{cal},
    \label{eq:modmatcal1}
\end{equation}
where the superscript $T$ represents the matrix transpose.

Since the modulation matrix ${\bf O}$ changes with the modulation cycle number, it
has to be determined for each and every modulation cycle. This can be done just by
continuing to record the signal for a number of modulation cycles continuously for a 
given known state of input polarization. The modulation matrix for the $k^{\mathrm{th}}$ 
modulation cycle is obtained from the corresponding  measured modulated intensities. 
In other words the Eq.~\ref{eq:modmatcal1} has to be solved for different
modulation cycles as follows.

\begin{equation}
    {\bf O}(k) = \left( {\bf I}^T_\mathrm{mod}(k) {\bf I}_\mathrm{mod}(k) \right)^{-1} {\bf I}^T_\mathrm{mod}(k) ~ {\bf S}_{cal}.
    \label{eq:modmatcal2}
\end{equation}
The corresponding demodulation for each modulation cycle is given by
\begin{equation}
    {\bf D(k)} = \left( {\bf O(k)}^T{\bf O}(k) \right)^{-1} {\bf O}^T(k).
    \label{eq:dmodmatcal2}
\end{equation}
With these calibration data the variation in modulation matrix from one modulation cycle to the other caused due to non-zero difference $\Delta\tau-\Delta t$ can be taken in to account.

Numerical experiments were carried out to check if the above proposed calibration method works. For this purpose a given modulation scheme for e.g. $n=8$ was chosen and calculated modulation matrices along with the corresponding demodulation matrices as a function of successive modulation cycles. The modulated intensity vectors of length $n$ were calculated for a given modulation scheme and for a given input Stokes vector as a function of successive modulation cycles.  In the next step the demodulation matrix of a given modulation cycle was used to retrieve the input Stokes vector from the corresponding modulated intensity vector. It was found that the input Stokes vector was fully retrieved.
However when an uncertainty in $dt-d\tau$ was introduced cross-talk among Stokes parameters was observed. 
Uncertainty in $d\tau-dt$ takes into account the instability in the slot time or the instability in the frame time or both. In order to introduce the uncertainty in $dt-d\tau$, random numbers were generated within a fixed range and the corresponding uncertainty in the angles was calculated by multiplying it with $\frac{2\pi}{T}$. 
This uncertainty in angle basically reflects in the uncertainty of $\theta_1$ and $\theta_2$ of  Eqs.~\ref{eq:modint2}~and~6 and hence the uncertainty in modulation and demodulation matrices. The same numerical experiments were carried out as explained in the beginning of this paragraph by inputting one Stokes parameter at a time. The above procedure was repeated for a few hundred times. It was found that with an uncertainty of $\pm$2~ms in $d\tau-dt$ the cross-talk among Stokes $Q$ and $U$ is in the order of 0.1\% which increases to 1\% if the the uncertainty is about 20~ms. Uncertainty in $dt-d\tau$ does not cause any cross-talk from Stokes~$I$ to $Q$, $U$ and $V$ and vice-versa. Another important cross-talk that is from Stokes $Q$ and $U$ to Stokes $V$ which is negligible. Even with 20~ms uncertainty the cross-talk is in the order of $10^{-5}$ or less. The effect of uncertainty in $d\tau-dt$ was checked for different modulation schemes and it was found that the cross-talk values are similar. This only depends on the magnitude of uncertainty in $d\tau-dt$.

\section{Conclusion}
It is found that the data volume can be optimized by choosing an appropriate  number of signal samples per half rotation than the usually adapted 8 samples.
Such a flexibility of choosing any number of samples exist with 
continuously rotating waveplate based modulator without significantly
changing the polarimetric efficiency.
The lowest number that can be chosen is 4 for full Stokes polarimetry.
However, use of QWP as the polarization modulator in the case of VELC/SP causes ambiguity in delineating the Stokes $I$ and $Q$ parameters and hence improper 
demodulation matrix with 4 samples per half rotation. 
But, this will not remain as an issue if the Stokes parameters are retrieved using the difference signal from the dual-beam.
The issue of varying modulation matrix from one modulation cycle to the next caused due to non-zero  difference between the retarder rotation period and the total time for recording the required samples is also
formulated in this paper. 
An effective solution is suggested to address the varying modulation matrix through polarimetric calibration without any need for hardware modification.

\begin{backmatter}

\bmsection{Acknowledgments}
We thank all the Scientists/Engineers at the various centres of ISRO such as URSC, SAC, LEOS, VSSC and Indian Institute of Astrophysics who have been contributing to Aditya-L1 mission. We gratefully acknowledge the ﬁnancial support from ISRO for this project.

\medskip

\bmsection{Disclosures} The authors declare no conflicts of interest.

\bigskip

\bmsection{Data availability} Data underlying the results presented in this paper are not publicly available at this time but may be obtained from the authors upon reasonable request.

\end{backmatter}

\bibliography{sample}

\begin{thebibliography}{10}
\newcommand{\enquote}[1]{``#1''}

\bibitem{Singh2011}
J.~{Singh}, B.~R. {Prasad}, P.~{Venkatakrishnan}, K.~{Sankarasubramanian},
  D.~{Banerjee}, R.~{Bayanna}, S.~{Mathew}, J.~{Murthy}, P.~{Subramanian},
  R.~{Ramesh}, S.~{Kathiravan}, S.~{Nagabhushana}, P.~K. {Mahesh}, P.~K.
  {Manoharan}, W.~{Uddin}, S.~{Sriram}, A.~{Kumar}, N.~{Srivastava}, K.~{Rao},
  C.~L. {Nagendra}, P.~{Chakraborthy}, K.~V. {Sriram}, R.~{Venkateswaran},
  T.~{Krishnamurthy}, P.~{Sreekumar}, K.~S. {Sarma}, R.~{Murthy}, K.~H.
  {Navalgund}, D.~R.~M. {Samudraiah}, P.~N. {Babu}, and A.~{Patra},
  \enquote{{Proposed visible emission line space solar coronagraph},}
  {\protect\JournalTitle{Current Science}} \textbf{100}, 167--174 (2011).

\bibitem{RaghavendraPrasad2017}
B.~{Raghavendra Prasad}, D.~{Banerjee}, J.~{Singh}, S.~{Nagabhushana},
  A.~{Kumar}, P.~U. {Kamath}, S.~{Kathiravan}, S.~{Venkata}, N.~{Rajkumar},
  V.~{Natarajan}, M.~{Juneja}, P.~{Somu}, V.~{Pant}, N.~{Shaji},
  K.~{Sankarsubramanian}, A.~{Patra}, R.~{Venkateswaran}, A.~A. {Adoni},
  S.~{Narendra}, T.~R. {Haridas}, S.~K. {Mathew}, R.~{Mohan Krishna},
  K.~{Amareswari}, and B.~{Jaiswal}, \enquote{{Visible Emission Line
  Coronagraph on Aditya-L1},} {\protect\JournalTitle{Current Science}}
  \textbf{113}, 613 (2017).

\bibitem{Rajkumar2018}
N.~{Raj Kumar}, B.~{Raghavendra Prasad}, J.~{Singh}, and S.~{Venkata},
  \enquote{{Optical design of visible emission line coronagraph on Indian space
  solar mission Aditya-L1},} {\protect\JournalTitle{Experimental Astronomy}}
  \textbf{45}, 219--229 (2018).

\bibitem{Donati90}
J.-F. {Donati}, M.~{Semel}, D.~E. {Rees}, K.~{Taylor}, and R.~D. {Robinson},
  \enquote{{Detection of a magnetic region on HR 1099},}
  {\protect\JournalTitle{Astron. Astrophys.}} \textbf{232}, L1--L4 (1990).

\bibitem{Semel93}
M.~{Semel}, J.-F. {Donati}, and D.~E. {Rees}, \enquote{{Zeeman-Doppler imaging
  of active stars. 3: Instrumental and technical considerations},}
  {\protect\JournalTitle{Astron. Astrophys}} \textbf{278}, 231--237 (1993).

\bibitem{Lites87}
B.~W. {Lites}, \enquote{{Rotating waveplates as polarization modulators for
  Stokes polarimetry of the sun - Evaluation of seeing-induced crosstalk
  errors},} {\protect\JournalTitle{Appl. Opt.}} \textbf{26}, 3838--3845 (1987).

\bibitem{Elmore92}
D.~F. {Elmore}, B.~W. {Lites}, S.~{Tomczyk}, A.~P. {Skumanich}, R.~B. {Dunn},
  J.~A. {Schuenke}, K.~V. {Streander}, T.~W. {Leach}, C.~W. {Chambellan}, and
  H.~K. {Hull}, \enquote{{The Advanced Stokes Polarimeter - A new instrument
  for solar magnetic field research},} in \emph{Polarization Analysis and
  Measurement,}  vol. 1746 of \emph{Society of Photo-Optical Instrumentation
  Engineers (SPIE) Conference Series} D.~H. {Goldstein} and R.~A. {Chipman},
  eds. (1992), pp. 22--33.

\bibitem{Socas-Navarro06}
H.~{Socas-Navarro}, D.~{Elmore}, A.~{Pietarila}, A.~{Darnell}, B.~W. {Lites},
  S.~{Tomczyk}, and S.~{Hegwer}, \enquote{{Spinor: Visible and Infrared
  Spectro-Polarimetry at the National Solar Observatory},}
  {\protect\JournalTitle{Solar Phys.}} \textbf{235}, 55--73 (2006).

\bibitem{Ishikawa2011}
R.~{Ishikawa}, T.~{Bando}, D.~{Fujimura}, H.~{Hara}, R.~{Kano}, T.~{Kobiki},
  N.~{Narukage}, S.~{Tsuneta}, K.~{Ueda}, H.~{Wantanabe}, K.~{Kobayashi},
  J.~{Trujillo Bueno}, R.~{Manso Sainz}, J.~{Stepan}, B.~{de Pontieu},
  M.~{Carlsson}, and R.~{Casini}, \enquote{{A Sounding Rocket Experiment for
  Spectropolarimetric Observations with the Ly$_{{\ensuremath{\alpha}}}$ Line
  at 121.6 nm (CLASP)},} in \emph{Solar Polarization 6,}  vol. 437 of
  \emph{Astronomical Society of the Pacific Conference Series} J.~R. {Kuhn},
  D.~M. {Harrington}, H.~{Lin}, S.~V. {Berdyugina}, J.~{Trujillo-Bueno}, S.~L.
  {Keil}, and T.~{Rimmele}, eds. (2011), p. 287.

\bibitem{Ichimoto2008}
K.~{Ichimoto}, B.~{Lites}, D.~{Elmore}, Y.~{Suematsu}, S.~{Tsuneta},
  Y.~{Katsukawa}, T.~{Shimizu}, R.~{Shine}, T.~{Tarbell}, A.~{Title},
  J.~{Kiyohara}, K.~{Shinoda}, G.~{Card}, A.~{Lecinski}, K.~{Streander},
  M.~{Nakagiri}, M.~{Miyashita}, M.~{Noguchi}, C.~{Hoffmann}, and T.~{Cruz},
  \enquote{{Polarization Calibration of the Solar Optical Telescope onboard
  Hinode},} {\protect\JournalTitle{Solar Phys.}} \textbf{249}, 233--261 (2008).

\bibitem{Lin2000}
H.~{Lin}, M.~J. {Penn}, and S.~{Tomczyk}, \enquote{{A New Precise Measurement
  of the Coronal Magnetic Field Strength},} {\protect\JournalTitle{Astrophys.
  J. Lett.}} \textbf{541}, L83--L86 (2000).

\bibitem{Lin2004}
H.~{Lin}, J.~R. {Kuhn}, and R.~{Coulter}, \enquote{{Coronal Magnetic Field
  Measurements},} {\protect\JournalTitle{Astrophys. J. Lett.}} \textbf{613},
  L177--L180 (2004).

\bibitem{Gibson2017}
S.~E. {Gibson}, K.~{Dalmasse}, L.~A. {Rachmeler}, M.~L. {De Rosa},
  S.~{Tomczyk}, G.~{de Toma}, J.~{Burkepile}, and M.~{Galloy},
  \enquote{{Magnetic Nulls and Super-radial Expansion in the Solar Corona},}
  {\protect\JournalTitle{Astrophys. J. Lett.}} \textbf{840}, L13 (2017).

\bibitem{Singh2019}
J.~{Singh}, B.~R. {Prasad}, S.~{Venkata}, and A.~{Kumar}, \enquote{{Exploring
  the outer emission corona spectroscopically by using Visible Emission Line
  Coronagraph (VELC) on board ADITYA-L1 mission},}
  {\protect\JournalTitle{Advances in Space Research}} \textbf{64}, 1455--1464
  (2019).

\bibitem{Liang2019}
Y.~{Liang}, Z.~{Qu}, Y.~{Zhong}, Z.~{Song}, and S.~{Li}, \enquote{{Analysis of
  errors in polarimetry using a rotating waveplate},}
  {\protect\JournalTitle{Appl. Opt.}} \textbf{58}, 9883 (2019).

\bibitem{Sahal-Brechot1977}
S.~{Sahal-Brechot}, \enquote{{Calculation of the polarization degree of the
  infrared lines of Fe XIII of the solar corona.}}
  {\protect\JournalTitle{Astrophys. J.}} \textbf{213}, 887--899 (1977).

\bibitem{LinCasini2000}
H.~{Lin} and R.~{Casini}, \enquote{{A Classical Theory of Coronal Emission Line
  Polarization},} {\protect\JournalTitle{Astrophys. J.}} \textbf{542}, 528--534
  (2000).

\bibitem{Singh1999}
J.~{Singh}, K.~{Ichimoto}, H.~{Imai}, T.~{Sakurai}, and A.~{Takeda},
  \enquote{{Spectroscopic Studies of the Solar Corona I. Spatial Variations in
  Line Parameters of Green and Red Coronal Lines},} {\protect\JournalTitle{Pub.
  Astron. Soc. Japan}} \textbf{51}, 269--276 (1999).

\bibitem{Singh2002}
J.~{Singh}, T.~{Sakurai}, K.~{Ichimoto}, and A.~{Takeda},
  \enquote{{Spectroscopic Studies of the Solar Corona III. Density Diagnostics
  Using the Infrared Lines of Fe XIII},} {\protect\JournalTitle{Pub. Astron.
  Soc. Japan}} \textbf{54}, 807--816 (2002).

\bibitem{Singh2003}
J.~{Singh}, K.~{Ichimoto}, T.~{Sakurai}, and S.~{Muneer},
  \enquote{{Spectroscopic Studies of the Solar Corona. IV. Physical Properties
  of Coronal Structure},} {\protect\JournalTitle{Astrophys. J.}} \textbf{585},
  516--523 (2003).

\bibitem{Tomczyk2008}
S.~{Tomczyk}, G.~L. {Card}, T.~{Darnell}, D.~F. {Elmore}, R.~{Lull}, P.~G.
  {Nelson}, K.~V. {Streander}, J.~{Burkepile}, R.~{Casini}, and P.~G. {Judge},
  \enquote{{An Instrument to Measure Coronal Emission Line Polarization},}
  {\protect\JournalTitle{Solar Phys.}} \textbf{247}, 411--428 (2008).

\bibitem{delToroIniesta00}
J.~C. {del Toro Iniesta} and M.~{Collados}, \enquote{{Optimum Modulation and
  Demodulation Matrices for Solar Polarimetry},} {\protect\JournalTitle{Appl.
  Opt.}} \textbf{39}, 1637--1642 (2000).

\bibitem{Collados07b}
M.~{Collados}, \enquote{{Polarimetry in the visible and near infrared},} in
  \emph{Modern solar facilities - advanced solar science,}  F.~{Kneer}, K.~G.
  {Puschmann}, and A.~D. {Wittmann}, eds. (2007), p. 143.

\bibitem{Ishikawa2014}
R.~{Ishikawa}, N.~{Narukage}, M.~{Kubo}, S.~{Ishikawa}, R.~{Kano}, and
  S.~{Tsuneta}, \enquote{{Strategy for Realizing High-Precision VUV
  Spectro-Polarimeter},} {\protect\JournalTitle{Solar Phys.}} \textbf{289},
  4727--4747 (2014).

\end{thebibliography}

\bibliographyfullrefs{sample}



\end{document}